June 10, 1994

# Is quantum mechanics compatible with an entirely deterministic universe?

*László E. Szabó*[*]

Center for Philosophy of Science
University of Pittsburgh

## Abstract

*It will be argued that 1) the Bell inequalities are not equivalent with those inequalities derived by Pitowsky and others that indicate the Kolmogorovity of a probability model, 2) the original Bell inequalities are irrelevant to both the question of whether or not quantum mechanics is a Kolmogorovian theory as well as the problem of determinism, whereas 3) the Pitowsky-type inequalities are not violated by quantum mechanics, hence 4) quantum mechanics is a Kolmogorovian probability theory, therefore, 5) it is compatible with an entirely deterministic universe.*

[*] On leave from the Institute for Theoretical Physics, Eötvös University, Budapest

# 1. Introduction

Is our world deterministic or indeterministic? Isn't everything already written in a Big (4-dimensional) Book? Is there a Becoming of future events, or is time merely subjective and Being timeless? Is there any difference between the ontological and epistemological probabilities? What about the Free Will? How many universes do we want? Do we need Branching Histories to describe our world? All these intriguing questions of philosophy of science are related to the very basic features of quantum mechanical probabilities. The debates about determinism/indeterminism are centered around the problem of whether or not there exist ontological modalities. Ontological modality means that *"at a given moment in the history of the world there are a variety of ways in which affairs might carry on. Before the toss of the coin there are two things that could happen, either Heads up or Tails up. This possibility is not merely epistemic, but <u>in re.</u>"* (Belnap & Green, forthcoming) The other possibility is that any stochasticity is merely epistemic, related to the lack of knowledge of the states of affairs.

Many believe that physics can provide some hints for solving some of the related philosophical problems. As Reichenbach (1956) writes with connection to a very close issue, *"There is no other way to solve the problem of time than... through physics. ... If time is objective the physicist must have discovered this fact, if there is Becoming the physicist must know it; but if time is merely subjective and Being is timeless, then physicist must have been able to ignore time in his construction of reality."* However, classical (statistical) physics leaves this philosophical problem unsolved, since the stochastic models of classical statistical physics are compatible with both the assumption of an underlying ontologically deterministic as well as indeterministic theory. Though, according to the common belief quantum mechanics is *not* compatible with a deterministic universe. Determinism, as it is used here, does not (necessarily) mean a functional relationship between earlier and later events, in the sense of computability, as determinism is usually understood in sciences (Cf. Grünbaum 1963 pp. 314-329, Earman 1986 and Faye 1989 Chapter 3). In other words, we do not assume that earlier time slices of the universe are determinately related, *by laws of nature*, to later time slices. Instead, determinism is used in a very deep, ontological sense. In Prior's branching time or Belnap's branching space-time terminology: there is no branching point and there is only one history (Belnap 1992). The assumption of ontological determinism does not contradict stochastic physical theories, which reflect the emergence of indeterminism at epistemic level. However, I want to emphasize that there is a testable physical consequence of the ontological determinism. Namely, the probabilities of the future events in a deterministic world should be interpreted as weighted averages of the possible truth value assignments. But we also know that this is true only if the corresponding probabilities admit a Kolmogorovian representation (Pitowsky 1989). According to widespread opinion, quantum mechanics is not compatible even with the assumption of the ontological determinism. As for instance Jeffrey Bub (1994) writes: *"... we know that we can't interpret the probabilities defined by the quantum state epistemically as measures over the different possible truth value assignments to all the propositions..."*.

In this paper, I challenge this conclusion by showing that no rigorous proof exists that the probabilities defined by quantum mechanics should be non-Kolmogorovian. Consequently, nothing implies that quantum theory is incompatible with a deterministic universe.

# 2. Preliminaries

According to the standard formulation, quantum mechanics can be regarded as a probability theory defined over the subspace lattice $L(H)$ of a Hilbert space *H*.

**Definition 1** *A non negative real function* $\mu$ *on* $L(H)$ *is called a* probability measure *if* $\mu(H) = 1$ *and if whenever* $E_1, E_2, \ldots E_n, \ldots$ *are pairwise orthogonal subspaces, and* $E = \bigvee_{i=1}^{\infty} E_i$ *then* $\mu(E) = \sum_{i=1}^{\infty} \mu(E_i).$

**Theorem 1 (Gleason)** *If H is a real or complex Hilbert space of dimension greater than 2, and* $\mu$ *is a probability measure on* $L(H)$, *then there exists a density operator W on H such that for each* $E \in L(H)$ *we have* $\mu(E) = tr(WE).$

And the converse also holds: for each density operator *W* there exists a probability measure $\mu_W$ such that for each $E \in L(H)$ we have $\mu_W(E) = tr(WE)$.

The *statistical algorithm* of quantum mechanics says:

$$p\left(\begin{array}{c}\text{the measured value of physical quantity } Q \\ \text{belongs to a Borel set } \omega \text{ of the spectrum of } Q\end{array}\right) = tr(WE_\omega)$$

where $E_\omega$ is the projector measure of the Borel set $\omega$. Since $E_\omega \in L(H)$, *p* can be regarded as the *restriction* of a $\mu_W$. And this fact justifies the above formulation of quantum mechanics.

**Definition 2** *A probability measure is* dispersion free *if* $\mu(E) \in \{0,1\}$ *for each* $E \in L(H)$.

**Definition 3** *A probabilistic theory is* deterministic *if each probability measure can be represented as a convex linear combination of dispersion free measures:*

$$\mu(A) = \sum_i \lambda_i \sigma_i(A) \qquad \sigma_i \text{ is dispersion free}$$
$$\sum_i \lambda_i = 1$$

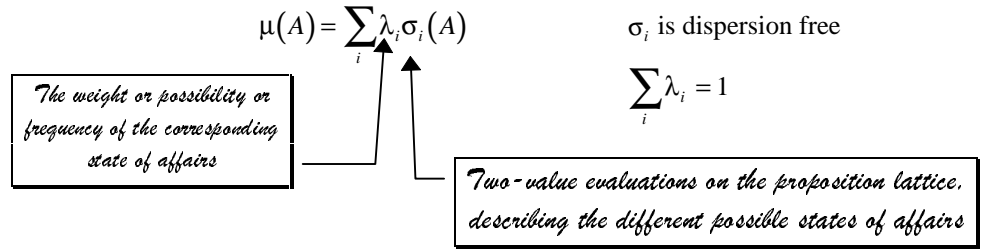

*The weight or possibility or frequency of the corresponding state of affairs*

*Two-value evaluations on the proposition lattice, describing the different possible states of affairs*

### 2.1. Von Neumann

The widespread conviction that we cannot interpret the probability defined by a quantum state as weighted average of the different possible truth value assignments to all the propositions goes back to J. von Neumann's first 1932 proof (Neumann 1955) of the non-existence of dispersion free states in quantum mechanics. His proof was based on four assumptions. The fourth, essential assumption is this: any real linear combination of Hermitian operators representing observables represents another observable, and the same linear combination of expectation values is the expectation value of the combination. But, this condition is one of the crucial weaknesses of Neumann's proof. The actual physical process that goes on during the measurement of the sum *A* + *B* is sometimes entirely different from the process for measuring *A* and *B* separately. Therefore, the additivity requirement in this assumption, which is satisfied by quantum states in the present theory, is a completely unjustified presupposition for the non-quantum-mechanical states in a future theory. And the fact that his presuppositions were so deeply rooted in the present Hilbert space quantum mechanics was a vulnerable point of von Neumann's proof, which has been intensively criticized in many of the succeeding publications on the problem of hidden variables (see Bohm 1957 and Bell 1966). Actually the same critique holds for Gleason's improved version of the Neumann theorem (see Clauser and Shimony 1978) as well as for a most recent proof of non-existence of hidden variables given by Greenberger, Horne, Shimony and Zeilinger (1990), which was summarized by Mermin (1990) in a very concise way (see Bohm and Hiley 1993).

## *2.2. Jauch and Piron*

An important contribution to understanding the nature of probabilities in quantum mechanics was made by Jauch and Piron (1963). They did not base their proof on the details of the formalism of quantum theory. Instead of $L(H)$ they regarded a more general orthocomplemented lattice of events $L$. On such a lattice one can define probability measures in the same way as above. They proved the following

**Theorem 2** *If L is deterministic then it is distributive.*

So, the question was reduced to this: *are there any non-compatible physical events/propositions?* They attempted to illustrate the non-distributivity of the lattice of physical propositions in an empirical way, independently of quantum mechanics. The suggested empirical illustration was this: A photon has property *A* if it passes a polar filter of angle $\alpha$ (Figure 1). Let *B* be the same property for angle $\beta$. The orthocomplement $A^\perp$ of *A*, corresponds to the same property with angle $\alpha+\pi/2$. If the property lattice is distributive, then for arbitrary *A* and *B* we have

$$(A \wedge B) \vee (A \wedge B^\perp) \vee (A^\perp \wedge B) \vee (A \wedge B^\perp) = I \qquad (1)$$

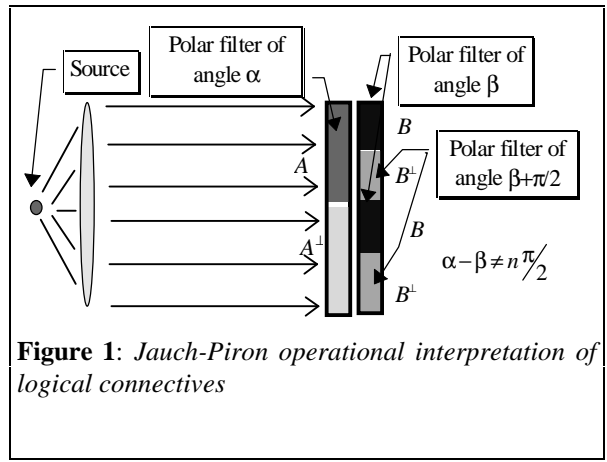

**Figure 1**: *Jauch-Piron operational interpretation of logical connectives*

The left hand side corresponds, they argued, to the experimental arrangement shown in Figure 1. The empirical violation of equation (1) is obvious: the corresponding system of filters should be completely transparent, but it is not. Consequently, the lattice of physical properties is not distributive; therefore, the probability measures defined on it cannot be deterministic. The advantage of the Jauch-Piron approach is that it uses only some very general assumptions about the lattice of physical properties. This proof of the non-existence of hidden parameters would be very convincing, if we didn't have to reject the above operational interpretation of logical connectives. The following difficulty arises. Let *A* be a property. According to Piron (1976), the corresponding event $\overline{A}$ is defined as the 'yes' outcome of a yes-no measurement testifying to the presence of property *A*. However, it is problematic whether the algebraic structure of *events* is the same as the algebraic structure of the *properties*. Namely, is it true that $\overline{A}\,\&\,\overline{B} = \overline{A \wedge B}$? I do not believe so[*]. As Bell (1987) rightly remarked (Cf. Forrest 1988 pp. 40-45), the Jauch-Piron operational interpretation of the "logical" connectives is misleading. The test of property $A \wedge B$ is a third measurement corresponding to entirely new physical circumstances. These circumstances sometimes have nothing to do with the experimental situation in which $A$ or $B$ is tested separately, especially if $A$ and $B$ are non-commuting properties.

### 2.3. Bell

The next historical step was Bell's analysis of the EPR experiment. Bell's approach to the problem of hidden variables had two advantages. 1) He avoided the problem of conjunction by referring only to conjunctions of outcomes belonging to commuting quantities. 2) Even though he used part of the machinery of quantum mechanics, one does not need to use it, but only elementary probability calculus and the experimental results. And that is why his proof of the non-existence of (local) hidden variables has been regarded as the most serious.

Consider an experiment corresponding to the Clauser and Horne derivation of the inequalities. It is like Aspect's experiment with spin-1/2 particles (Figure 2). Briefly recall the usual assumptions describing the intuitive notion of a local hidden variable. Assume that there is a parameter $\lambda$ taken as an element of a probability space $\langle \Lambda, \Sigma(\Lambda), \rho \rangle$, such that the quantum mechanical probabilities can be represented as follows:

$$p(\lambda, A \wedge B)_{a,b} = p(\lambda, A)_a \, p(\lambda, B)_b$$
$$p(A)_a = \int_\Lambda p(\lambda, A)_a \, d\rho$$
$$p(B)_b = \int_\Lambda p(\lambda, B)_b \, d\rho \qquad (2)$$
$$p(A \wedge B)_{a,b} = \int_\Lambda p(\lambda, A)_a \, p(\lambda, B)_b \, d\rho$$

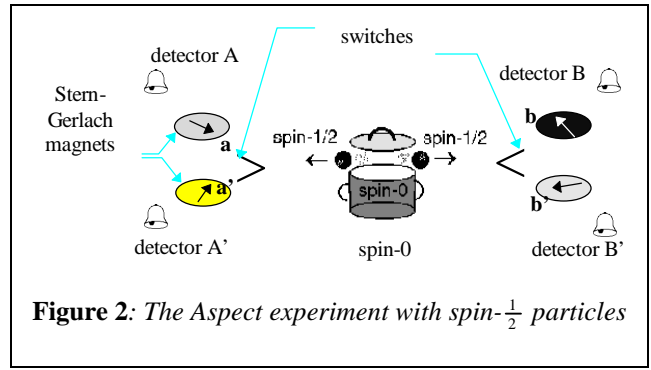

**Figure 2**: *The Aspect experiment with spin-$\frac{1}{2}$ particles*

For real numbers such that

$$0 \leq x, x', y, y' \leq 1$$

the following elementary inequality holds

$$-1 \leq xy - xy' + x'y + x'y' - x' - y \leq 0$$

Applying this inequality, we have

---

[*] This question is closely related to one of the main difficulties of quantum logic: how to interpret the "conjunctions" in case of non-commuting projectors. I plan to discuss this problem elsewhere.

$$-1 \leq p(\lambda, A \wedge B)_{a,b} - p(\lambda, A \wedge B')_{a,b'} + p(\lambda, A' \wedge B)_{a',b} + p(\lambda, A' \wedge B')_{a',b'} - p(\lambda, A')_{a'} - p(\lambda, B)_{b} \leq 0$$

Integrating this inequality we have

$$-1 \leq p(A \wedge B)_{a,b} - p(A \wedge B')_{a,b'} + p(A' \wedge B)_{a',b} + p(A' \wedge B')_{a',b'} - p(A')_{a'} - p(B)_{b} \leq 0 \qquad (3)$$

This is one of the well-known Clauser-Horne inequalities (one can get all the others by varying the roles of $A, A', B, B'$). Returning to the Aspect experiment, consider the following events:

$A$      left electron has spin "up" in direction **a**
$A'$      left electron has spin "up" in direction **a**$'$
$B$      right electron has spin "up" in direction **b**
$B'$      right electron has spin "up" in direction **b**$'$

As is well known, in case $\angle(\mathbf{x},\mathbf{y}) = \angle(\mathbf{y},\mathbf{w}) = \angle(\mathbf{x},\mathbf{w}) = 120°$ and $\mathbf{x} = \mathbf{z}$, the quantum mechanical predictions as well as the experimental results are

$$p(A)_a = p(A')_{a'} = p(B)_b = p(B')_{b'} = \tfrac{1}{2}$$
$$p(A \wedge B)_{a,b} = p(A \wedge B')_{a,b'} = p(A' \wedge B')_{a',b'} = \tfrac{3}{8} \qquad (4)$$
$$p(A' \wedge B)_{a',b} = 0$$

These probabilities violate the Clauser-Horne inequality. Thus, according to the usual conclusion, there is no local hidden variable theory reproducing the quantum mechanical probabilities.

## *2.4. Pitowsky*

In the above derivation of Clauser-Horne inequalities, as in any other derivation of Bell-type inequalities, *"... the physical aspects of the problem are intermingled with the purely mathematical character of the derivation of the inequalities"*, says Pitowsky (1989, p. 49). And he continues: *"This is a source of prevailing confusion, as if Bell-type inequalities have, in themselves something to do with physics. But they do not." "... these inequalities follow directly from the theory of probability or, if you like, from propositional logic. It is only their violation by quantum frequencies which makes them important for the foundations of physics."* In the rest of this paper I challenge this conclusion at two different points. 1) I will show that Pitowsky's inequalities, derived from the Kolmogorovian probability theory, are not the same as the Bell inequalities. 2) Pitowsky's inequalities are not violated by quantum mechanics. But, first, recall Pitowsky's important theorem about the conditions under which a probability theory is Kolmogorovian.

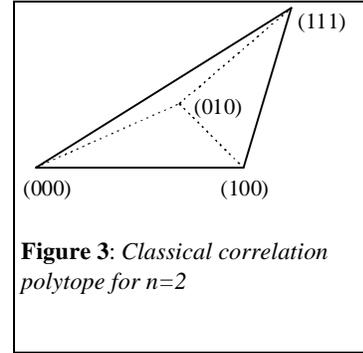

**Figure 3**: *Classical correlation polytope for n=2*

Let $S$ be a set of pairs of integers $S \subseteq \{\{i,j\} \mid 1 \leq i \leq j \leq n\}$. Denote by $R(n,S)$ the linear space of real vectors having a form like $(f_1 f_2 \ldots f_n \ldots f_{ij} \ldots)$. For each $\varepsilon \in \{0,1\}^n$, let $u^\varepsilon$ be the following vector in $R(n,S)$:

$$u_i^\varepsilon = \varepsilon_i \qquad 1 \leq i \leq n$$
$$u_{ij}^\varepsilon = \varepsilon_i \varepsilon_j \quad \{i,j\} \in S \qquad (5)$$

The classical correlation polytope $C(n,S)$ is the closed convex hull in $R(n,S)$ of vectors $\{u^\varepsilon\}_{\varepsilon \in \{0,1\}^n}$:

$$C(n,S) := \left\{ a \in R(n,S) \mid a = \sum_{z \in \{0,1\}^n} \lambda_z u^z \text{ such that } \lambda_z \geq 0 \text{ and } \sum \lambda_z = 1 \right\} \tag{6}$$

Let $\mathbf{p} = (p_1, \ldots p_n, \ldots p_{ij}, \ldots) \in R(n,S)$. We will then say that $\mathbf{p}$ *has a Kolmogorovian representation* if there exist a Kolmogorovian probability space $(\Omega, \Sigma, \mu)$ and events $A_1, A_2, \ldots A_n \in \Sigma$ such that

$$p_i = \mu(A_i) \qquad 1 \leq i \leq n$$
$$p_{ij} = \mu(A_i \cap A_j) \quad \{i,j\} \in S$$

**Theorem 3 (Pitowsky, 1989)** *A correlation vector* $\mathbf{p} = (p_1 p_2 \ldots p_n \ldots p_{ij} \ldots)$ *has a Kolmogorovian representation if and only if* $\mathbf{p} \in C(n,S)$.

From the definition of the polytope, equations (5) and (6), it is obvious that the condition $\mathbf{p} \in C(n,S)$ has the following meaning: the probabilities can be represented as weighted averages of the classical truth values.

In case $n = 4$ and $S = S_4 = \{\{1,3\}, \{1,4\}, \{2,3\}, \{2,4\}\}$, the condition $\mathbf{p} \in C(n,S)$ is equivalent with the following inequalities:

$$0 \leq p(A_i \wedge A_j) \leq p(A_i) \leq 1$$
$$0 \leq p(A_i \wedge A_j) \leq p(A_j) \leq 1 \qquad i = 1,2 \qquad j = 3,4$$
$$p(A_i) + p(A_j) - p(A_i \wedge A_j) \leq 1$$
$$-1 \leq p(A_1 \wedge A_3) + p(A_1 \wedge A_4) + p(A_2 \wedge A_4) - p(A_2 \wedge A_3) - p(A_1) - p(A_4) \leq 0$$
$$-1 \leq p(A_2 \wedge A_3) + p(A_2 \wedge A_4) + p(A_1 \wedge A_4) - p(A_1 \wedge A_3) - p(A_2) - p(A_4) \leq 0 \tag{7}$$
$$-1 \leq p(A_1 \wedge A_4) + p(A_1 \wedge A_3) + p(A_2 \wedge A_3) - p(A_2 \wedge A_4) - p(A_1) - p(A_3) \leq 0$$
$$-1 \leq p(A_2 \wedge A_4) + p(A_2 \wedge A_3) + p(A_1 \wedge A_3) - p(A_1 \wedge A_4) - p(A_2) - p(A_3) \leq 0$$

The last inequality of (7) really reminds one of the inequality (3) if $A_1 = A$, $A_2 = A'$, $A_3 = B$ and $A_4 = B'$. This explains why Pitowsky calls inequalities (7) "the Clauser-Horne inequalities", and the corresponding correlation polytope $C(4,S)$ "the Clauser-Horne polytope". Therefore, substituting for the probabilities in the last inequality of (7) the same values as were calculated from quantum mechanics in (4), we have

$$\frac{3}{8} + \frac{3}{8} + \frac{3}{8} - 0 - \frac{1}{2} - \frac{1}{2} = \frac{1}{8} > 0 \tag{8}$$

Consequently,

$$\mathbf{p} = \left( \frac{1}{2} \frac{1}{2} \frac{1}{2} \frac{1}{2} \frac{3}{8} \frac{3}{8} 0 \frac{3}{8} \right) \notin C(4,S)$$

Now, Pitowsky concludes: *"We have demonstrated that $\mathbf{p} \notin C(n,S)$ and therefore we cannot explain the statistical outcome by assuming that the source is an "urn", containing electron pairs in the singlet state, such that the distribution of the properties A, A', B, B' in this "urn" is fixed before the measurement."* (Pitowsky, p. 82; notation changed for sake of uniformity). *"The violations of these constraints on correlations by quantum frequencies thus poses a major problem for all schools of classical probability. I take this fact to be the major source of difficulty which underlies the interpretation of quantum theory."* (Ibid., p. 87).

# 3. The non-equality of the Bell-inequalities

At this essential point I challenge the above conclusion of Pitowsky: in formula (8) we substituted *incorrect values* for the probabilities of inequality (7). In order to see why, we have to investigate the exact meaning of the probabilities in quantum mechanics and in Pitowsky's inequalities.

Whenever we talk about probabilities *we have to specify the system of conditions under which the probabilities are understood.* Sometimes these conditions are given only tacitly, but without such conditions the probabilities are meaningless. The meaning of the quantum mechanical probability, $tr(WA)$, is this: "the probability of getting outcome $A$ *given* that the measurement $a$ is performed". Let $p(A)_a$ denote such a conditional probability. Throughout the derivation of the original Clauser-Horne inequalities (3) we used conditional probabilities of this type[*]:

$$-1 \le p(A \wedge B)_{a,b} - p(A \wedge B')_{a,b} + p(A' \wedge B)_{a',b} + p(A' \wedge B')_{a',b'} - p(A')_{a'} - p(B)_b \le 0$$

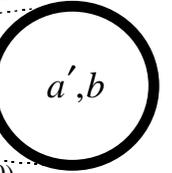

Making the tacit conditions explicit we have

$$-1 \le p(A \wedge B)_{a,b} - p(A \wedge B')_{a,b'} + p(A' \wedge B)_{a',b} + p(A' \wedge B')_{a',b'} - p(A')_{a'} - p(B)_b \le 0 \quad (10)$$

It is remarkable that in the original Clauser-Horne inequality (10) we have probabilities corresponding to *different* conditions.

At the same time the probabilities in the Pitowsky theorem belong to *one common set of conditions*, as a consequence of the fact that the probability measure on any Kolmogorovian probability space also belongs to one common set of conditions. These common conditions can be formulated in various ways. In case of the EPR experiment they can be the prepared physical circumstances at the moment of decay of the source particle. For brevity this primary common set of conditions is not indicated in our future notations. Since $A$ entails $a$, one can, of course, calculate the "unconditioned" probabilities in the usual way:

$$p(A) = tr(WA) \cdot p(a - \text{measurement}) \quad (11)$$

Actually these are the probabilities that one can measure in a laboratory by counting the "beeps" given by a detector responsible for outcome $A$. Assume that the switches (Figure 2) choose between the corresponding two measurements with equal probability: $p(a) = p(a') = p(b) = p(b') = \frac{1}{2}$. So, the correct numbers in (8) should be:

$$\frac{3}{32} + \frac{3}{32} + \frac{3}{32} - 0 - \frac{1}{4} - \frac{1}{4} = -\frac{7}{32} \in [-1, 0] \quad (12)$$

Consequently,

$$\mathbf{p} = \left(\frac{1}{4}\frac{1}{4}\frac{1}{4}\frac{1}{4}\frac{3}{32}\frac{3}{32} 0 \frac{3}{32}\right) \in C(4, S_4) \quad (13)$$

This means that the *measured probabilities* in the EPR experiment *are Kolmogorovian*!

Thus, the original Bell (Clauser-Horne) inequalities, such as (10), which contain conditional probabilities that belong to different conditions, *are not the same inequalities* as Pitowsky's "Bell (Clauser-Horne)" inequalities like (7).

---

[*] I must apologize if I made trouble for the reader by playing with nearly invisible subscripts throughout section 2.3. I only aimed to symbolize that those conditions are always there, but neglected by people.

# 4. Is quantum mechanics compatible with a deterministic universe?

The fact that Pitowsky's inequalities are not violated leads to the conclusion that *the EPR experiments do not provide empirical evidence against the Kolmogorovia character of quantum probabilities*. This conclusion is entirely contrary to that of Pitowsky himself[*]. One can ask now several questions: 1) What is the conceptual origin of Pitowsky's argumentation? 2) What is the deeper meaning of the fact that the EPR experiments do not violate Pitowsky's version of "Clauser-Horne" inequalities, while they do violate the original Clauser-Horne inequalities? 3) If, as I suggest, quantum mechanics is a Kolmogorovian probability theory, what kind of consequences follow from this fact with respect to the hidden variable theories and the possibility of a deterministic explanation of quantum phenomenon?

## 4.1. Presumptions versus reality

In order to see the conceptual mistake in substituting the value of conditional probability $p(A)_a$ for the "unconditioned" probability $p(A)$, consider two distinct understandings of quantum mechanical probabilities:

a) $p(A)_a = tr(WA)$ is the probability of the occurrence of the outcome $A$ given that the measurement $a$ is performed.

b) $p(A)_a = tr(WA)$ is the probability that the system has the *property* that "the outcome $A$ occurs whenever measurement $a$ is performed".

If we accept version a) then the substitution $p(A)_a$ for $p(A)$ is a simple mistake of calculation. To assume b) leads to the following contradiction. In the case of the EPR experiment,

i.) according to interpretation b) one assumes that

> "the source is an 'urn', containing electron pairs in singleton state such that the distribution of the properties *A, A', B, B'* in this 'urn' is fixed before the measurement"

ii.) in this case one can correctly substitute $p(A)_a$ for $p(A)$, and have

$$\mathbf{p} \notin C(n, S)$$

*Contradiction!*

iii.) from which it follows that

> "we cannot explain the statistical outcome by assuming that the source is an 'urn', containing electron pairs in singleton state such that the distribution of the properties *A, A', B, B'* in this 'urn' is fixed before the measurement"

From this contradiction it does not at all follow that quantum mechanics is a non-Kolmogorovian probability theory. But, it does follow that b) is an untenable interpretation of the quantum mechanical probabilities.

---

[*] It also contradicts to some parts of my two earlier papers (1993, 1994) in which I uncritically recalled the conclusions of Pitowsky.

## *4.2. Violation of the original inequalities*

One can ask then, what does the violation of the original Bell-type inequalities mean, if does not mean that quantum mechanics is a non-Kolmogorovian theory? Compare the two kinds of Bell inequalities. Taking into account (11), the original Clauser-Horne inequalities can be expressed in the following way:

$$-1 \leq \frac{p(A \wedge B)}{p(a \wedge b)} - \frac{p(A \wedge B')}{p(a \wedge b')} + \frac{p(A' \wedge B)}{p(a' \wedge b)} + \frac{p(A' \wedge B')}{p(a' \wedge b')} - \frac{p(A')}{p(a')} - \frac{p(B)}{p(b)} \leq 0 \qquad (14)$$

The corresponding Pitowsky inequality is the last inequality of (7):

$$-1 \leq p(A \wedge B) - p(A \wedge B') + p(A' \wedge B) + p(A' \wedge B') - p(A') - p(B) \leq 0 \qquad (15)$$

However, inequalities (14) and (15) are entirely different. And there is no easy way to specify when the violation of one of them involves the violation of the other. But fortunately we do not need to deal with this problem since we can entirely neglect the original Bell inequalities. We can do that for two different reasons. First, as we have seen, they are completely irrelevant to the question of whether or not quantum mechanics is a Kolmogorovian theory. Second, they are irrelevant from the point of view of a common cause explanation of quantum correlations. And indeed, according to the understanding of quantum mechanical probabilities mentioned as version a) in section 4.1, the question of whether or not there exists such a hidden parameter that satisfies conditions

$$p(\lambda, A \wedge B)_{a,b} = p(\lambda, A)_a \, p(\lambda, B)_b$$
$$p(A)_a = \int_\Lambda p(\lambda, A)_a \, d\rho$$
$$p(B)_b = \int_\Lambda p(\lambda, B)_b \, d\rho \qquad (16)$$
$$p(A \wedge B)_{a,b} = \int_\Lambda p(\lambda, A)_a \, p(\lambda, B)_b \, d\rho$$

is meaningless. As a matter of fact, these conditions would express the intuitive conception of a common cause mechanism responsible for the EPR correlations, only if the probabilities in (16) were interpreted as the probabilities of appearances of "properties" of the system. This interpretation is nothing else but the version b) in section 4.1. Only in this case, we could infer from the violation of the original Clauser-Horne inequality to the violation of Einstein-locality. But, we have rejected version b) as a contradictory interpretation. In other words, since there are no "properties" corresponding to outcomes of measurements we can perform on the system (at least quantum mechanics has nothing to do with such properties), we do not need to explain "the correlation between spatially separated occurrence of such (non-existing) properties". But, there do exist (observable physical) events corresponding to performance-preparations of various measurements and other events which correspond to the outcomes. Each such event occurs with certain probability. What we observe in the EPR experiment is a correlation between spatially separated *outcomes*. And the question is whether or not a local hidden variable explanation is possible for such a correlation. Thus, the whole formulation of the common cause problem has to be reconsidered.

## 4.3. Correct formulation of the local hidden variable problem

A local hidden variable theory (considered as a mathematically well formulated representation of a deterministic and non-violating Einstein causality universe) has to reproduce the probabilities of the out-comes and that of the performance-preparation. The assumed "parameter" $\lambda \in \Lambda$ should represent the state of the part of the universe which belongs to the common past of the two separated measurements (**Figure 4**) such that for both, the outcomes and the performance-preparations we have

$$p(A) = \int_\Lambda p(\lambda, A) d\rho \qquad (17/1)$$

$$p(A') = \int_\Lambda p(\lambda, A') d\rho \qquad (17/2)$$

$$p(B) = \int_\Lambda p(\lambda, B) d\rho \qquad (17/3)$$

$$p(B') = \int_\Lambda p(\lambda, B') d\rho \qquad (17/4)$$

$$p(a) = \int_\Lambda p(\lambda, a) d\rho \qquad (17/5)$$

$$p(a') = \int_\Lambda p(\lambda, a') d\rho \qquad (17/6)$$

$$p(b) = \int_\Lambda p(\lambda, b) d\rho \qquad (17/7)$$

$$p(b') = \int_\Lambda p(\lambda, b') d\rho \qquad (17/8)$$

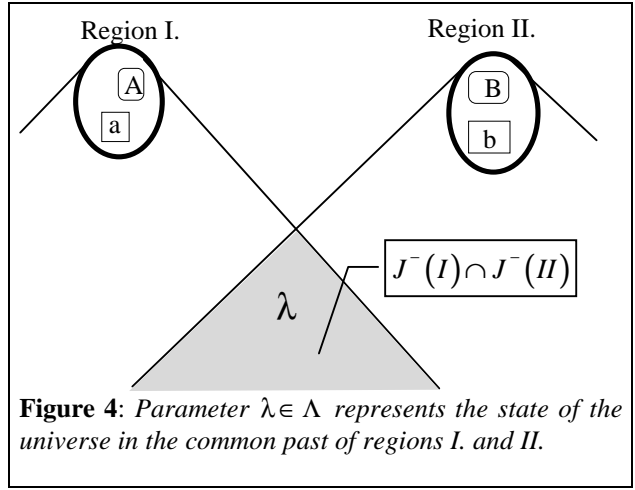

**Figure 4**: *Parameter $\lambda \in \Lambda$ represents the state of the universe in the common past of regions I. and II.*

The same holds for the conjunctions

$$p(A \wedge B) = \int_\Lambda p(\lambda, A \wedge B) d\rho \qquad (17/9)$$

$$p(A \wedge B') = \int_\Lambda p(\lambda, A \wedge B') d\rho \qquad (17/10)$$

$$\vdots$$

$$p(a' \wedge B') = \int_\Lambda p(\lambda, a' \wedge B') d\rho \qquad (17/24)$$

And, we also assume that the underlying hidden variable theory is Einstein-local. This means that the correlation between any two spatially separated events should be the consequence of the $\lambda$-dependence of the corresponding probabilities and not the consequence of a direct physical interaction. We can formulate this assumption via the following relations:

$$p(\lambda, A \wedge B) = p(\lambda, A) \cdot p(\lambda, B) \tag{18/1}$$
$$p(\lambda, A \wedge B') = p(\lambda, A) \cdot p(\lambda, B') \tag{18/2}$$
$$p(\lambda, A' \wedge B) = p(\lambda, A') \cdot p(\lambda, B) \tag{18/3}$$
$$p(\lambda, A' \wedge B') = p(\lambda, A') \cdot p(\lambda, B') \tag{18/4}$$
$$p(\lambda, a \wedge b) = p(\lambda, a) \cdot p(\lambda, b) \tag{18/5}$$
$$p(\lambda, a \wedge b') = p(\lambda, a) \cdot p(\lambda, b') \tag{18/6}$$
$$p(\lambda, a' \wedge b) = p(\lambda, a') \cdot p(\lambda, b) \tag{18/7}$$
$$p(\lambda, a' \wedge b') = p(\lambda, a') \cdot p(\lambda, b') \tag{18/8}$$
$$p(\lambda, A \wedge b) = p(\lambda, A) \cdot p(\lambda, b) \tag{18/9}$$
$$p(\lambda, A \wedge b') = p(\lambda, A) \cdot p(\lambda, b') \tag{18/10}$$
$$p(\lambda, A' \wedge b) = p(\lambda, A') \cdot p(\lambda, b) \tag{18/11}$$
$$p(\lambda, A' \wedge b') = p(\lambda, A') \cdot p(\lambda, b') \tag{18/12}$$
$$p(\lambda, a \wedge B) = p(\lambda, a) \cdot p(\lambda, B) \tag{18/13}$$
$$p(\lambda, a \wedge B') = p(\lambda, a) \cdot p(\lambda, B') \tag{18/14}$$
$$p(\lambda, a' \wedge B) = p(\lambda, a') \cdot p(\lambda, B) \tag{18/15}$$
$$p(\lambda, a' \wedge B') = p(\lambda, a') \cdot p(\lambda, B') \tag{18/16}$$

Relations (18/1-4) express the $\lambda$-level independence of the outcomes (in other words, the screening off the outcomes by the hidden parameter), relations (18/5-8) express the $\lambda$-level independence of the choices which measurement will be performed. Finally, equations (18/9-16) represent the $\lambda$-level independence of the outcomes from the spatially separated choices (required parameter independence).

Now, the question is whether there exists such a parameter satisfying conditions (17/1-24) and (18/1-16). One can use[*] the following

**Theorem 4** *With the notation of the section 2.4 consider events $A_1, \ldots A_n$ and a set of indexes S. Assume that a correlation vector $\mathbf{p} = \left(p(A_1) \ldots p(A_n) \ldots p(A_i \wedge A_j) \ldots\right)$ can be represented as convex combination of parameter-depending correlation vectors $\left(\pi(\lambda, A_1) \ldots \pi(\lambda, A_n) \ldots \pi(\lambda, A_i \wedge A_j) \ldots\right)$,*

$$p(A_i) = \int_\Lambda \pi(\lambda, A_i) d\rho(\lambda) \qquad \text{for } 1 \leq i \leq n$$
$$p(A_i \wedge A_j) = \int_\Lambda \pi(\lambda, A_i \wedge A_j) d\rho(\lambda) \qquad \text{for } \{i, j\} \in S \tag{19}$$

*such that*

$$\pi(\lambda, A_i \wedge A_j) = \pi(\lambda, A_i) \cdot \pi(\lambda, A_j) \qquad \text{for each } \{i, j\} \in S \tag{20}$$

*Then $\mathbf{p} \in C(n, S)$.*

(See Szabó 1993 for the proof.)
Equations (17/1-24) correspond to (19) and (18/1-16) to (20) in case

---

[*] We cannot use the formal repetion of the Clauser-Horne derivation described in section 2.3 because the number of events and the number of investigated conjunctions are larger then four.

$$A_1 = A \qquad A_5 = a$$
$$A_2 = A' \qquad A_6 = a'$$
$$A_3 = B \qquad A_7 = b$$
$$A_4 = B' \qquad A_8 = b'$$

$$S = S_{16} = \{\{1,3\},\{1,4\},\{2,3\},\{2,4\},\{5,7\},\{5,8\},\{6,7\},\{6,8\},\{1,7\},\{1,8\},\{2,7\},\{2,8\},\{5,3\},\{5,4\},\{6,3\},\{6,4\}\}$$

According to this theorem, if there exists a hidden parameter theory satisfying conditions (17/1-24) and (18/1-16) then the *observed* probabilities should satisfy condition $\mathbf{p} \in C(8, S_{16})$.

Let us gather all the information we know about the observed probabilities in the Aspect-type spin-correlation experiment. According to the section 3 we have

$$p(A) = p(A') = p(B) = p(B') = \frac{1}{4} \qquad (21/1)$$

$$p(a) = p(a') = p(b) = p(b') = \frac{1}{2} \qquad (21/2)$$

$$p(A \wedge B) = p(A \wedge B') = p(A' \wedge B') = \frac{3}{32} \qquad (21/3)$$

$$p(A' \wedge B) = 0 \qquad (21/4)$$

The probabilities with which the different measurements are chosen satisfy the following independence conditions:

$$p(a \wedge b) = p(a) \cdot p(b)$$
$$p(a \wedge b') = p(a) \cdot p(b')$$
$$p(a' \wedge b) = p(a') \cdot p(b)$$
$$p(a' \wedge b') = p(a') \cdot p(b')$$

Therefore

$$p(a \wedge b) = p(a \wedge b') = p(a' \wedge b) = p(a' \wedge b') = \frac{1}{4} \qquad (21/5)$$

Similarly, it is an observed fact that the outcomes on the one side are independent of the choices on the other side (observed parameter independence). This means that

$$p(A \wedge b) = p(A) \cdot p(b) \qquad\qquad p(a \wedge B) = p(a) \cdot p(B)$$
$$p(A \wedge b') = p(A) \cdot p(b') \qquad\qquad p(a \wedge B') = p(a) \cdot p(B')$$
$$p(A' \wedge b) = p(A') \cdot p(b) \qquad\qquad p(a' \wedge B) = p(a') \cdot p(B)$$
$$p(A' \wedge b') = p(A') \cdot p(b') \qquad\qquad p(a' \wedge B') = p(a') \cdot p(B')$$

According to these independence relations

$$p(A \wedge b) = p(A \wedge b') = p(A' \wedge b) = p(A' \wedge b') =$$
$$p(a \wedge B) = p(a \wedge B') = p(a' \wedge B) = p(a' \wedge B') = \frac{1}{8} \qquad (21/6)$$

We can collect the data from (21/1-6) in a correlation vector. The question is whether or not this correlation vector is contained in the classical correlation polytope:

$$\mathbf{p} = \left( \frac{1}{4} \frac{1}{4} \frac{1}{4} \frac{1}{4} \frac{1}{2} \frac{1}{2} \frac{1}{2} \frac{1}{2} \frac{3}{32} \frac{3}{32} \; 0 \; \frac{3}{32} \frac{1}{4} \frac{1}{4} \frac{1}{4} \frac{1}{4} \frac{1}{8} \frac{1}{8} \frac{1}{8} \frac{1}{8} \frac{1}{8} \frac{1}{8} \frac{1}{8} \frac{1}{8} \right) \stackrel{?}{\in} C(8, S_{16}) \quad (22)$$

In case $n > 4$ there are no inequalities derived which would be equivalent to the condition $\mathbf{p} \in C(n, S)$ (see Pitowsky 1989 for the details). We thus have to examine the geometric condition (22) directly. I testified condition (22) by computer and the result is affirmative:

$$\mathbf{p} = \left( \frac{1}{4} \frac{1}{4} \frac{1}{4} \frac{1}{4} \frac{1}{2} \frac{1}{2} \frac{1}{2} \frac{1}{2} \frac{3}{32} \frac{3}{32} \; 0 \; \frac{3}{32} \frac{1}{4} \frac{1}{4} \frac{1}{4} \frac{1}{4} \frac{1}{8} \frac{1}{8} \frac{1}{8} \frac{1}{8} \frac{1}{8} \frac{1}{8} \frac{1}{8} \frac{1}{8} \right) \in C(8, S_{16})$$

Consequently, *there is no proved disagreement between the assumptions* (17/1-24) *and* (18/1-16) *about a local hidden variable theory and the observations.* In other words, *the existence of a local hidden variable theory is not excluded.*

# 5. Conclusions

From my historical review it turned out that the Bell analysis of the EPR experiment was considered to be a clear and indeed the best illustration of the alleged fact that there is no (at least local) hidden variable theory which can reproduce the quantum mechanical probabilities. We recalled Pitowsky's theorem about the conditions under which a probability theory is Kolmogorovian. It is widely believed that Pitowsky's conditions are, as he claimed himself, equivalent with the Bell inequalities. Since quantum probabilities violate these inequalities, it is believed that quantum mechanics is not only incompatible with a local hidden parameter theory, but also incompatible with a Kolmogorovian probability-theoretic description. As a consequence of the non-Kolmogorovity, it is also sustained that quantum theory is incompatible with a ontologically deterministic world, i.e., the probabilities defined by the quantum states cannot be interpreted epistemically as weighted averages of the different possible truth value assignments to all the propositions. However, we have shown the following:

1. The Bell inequalities are not equivalent with those inequalities derived by Pitowsky.
2. The original Bell inequalities are irrelevant, while
3. the Pitowsky-type inequalities are not violated by quantum mechanics.
4. Quantum mechanics is a Kolmogorovian probability theory.
5. A more correct formulation of the hidden variable problem can be given.
6. There is no rigorous indication that quantum mechanics is not compatible with an entirely deterministic universe.

# Acknowledgements

I am extremely grateful to Professor Nuel Belnap for numerous discussions and for a careful reading of a draft version of this paper. For the latter task I also extend warm thanks to Jan Faye. For the kind hospitality in Pittsburgh I am indebted to the Center for Philosophy of Science where the research for this paper has been made. The financial support has been provided by the OTKA Foundation (Grant No. 1826), and by a fellowship from the Fulbright Foundation.